\documentstyle[12pt,aps]{revtex}

\newcommand{\bra}[1]{\mbox{$\langle #1 |$}}
\newcommand{\ket}[1]{\mbox{$| #1 \rangle$}}
\newcommand{\bracket}[2]{\mbox{$\langle {{#1}} \mathrel{ | {\vphantom
        {{#1} {#2}}} \kern-\nulldelimiterspace} {{#2}} \rangle$}}

\def\R{\hbox{$\mit I$\kern-.277em$\mit R$}}
\def\N{\hbox{$\mit I$\kern-.277em$\mit N$}}
\def\C{\hbox{$\mit I$\kern-.7em$\mit C$}}
\def\un{\leavevmode\hbox{\normalsize1\kern-4.6pt\large1}}

\begin{document}

\title{Quantum Correlations, Local Interactions and Error Correction}
\author{V. Vedral, M.A.Rippin, M.B.Plenio \\ Optics Section, \\ Blackett Laboratory, 
\\ Imperial College, \\ London SW7 2BZ, \\ England.}
\date{\today}
\maketitle

\begin{abstract}
We consider the effects of local interactions upon 
quantum mechanically entangled systems.  In particular we demonstrate 
that non-local correlations {\it cannot increase} through local 
operations on any of the subsystems, but that through the use of 
quantum error correction methods, correlations can be {\it maintained}. 
We provide two mathematical proofs that local general measurements 
cannot increase correlations, and also derive general conditions for 
quantum error correcting codes. Using these we show that local 
quantum error correction can preserve nonlocal features of
entangled quantum systems. We also demonstrate these results by 
use of specific examples employing correlated optical cavities 
interacting locally with resonant atoms.  By way of counter example, 
we also describe a mechanism by which correlations can be increased, 
which demonstrates the need for {\it non-local} interactions.  

\end{abstract}

PACS number(s) 03.65

\section{Introduction}

In the last year or so, much of the interest in quantum information
theory has been directed towards two related subjects: firstly in analysing so 
called purification procedures \cite{Gisin,Deutsch} and secondly in 
exploring the idea  of quantum error correction \cite{Steane1,Steane2,Calderbank1,Ekert1,Laflamme1,Knill1,Plenio1}, 
as well as examining the connections between the two \cite{Bennett1}. 
Purification procedures are based on Gisin's original proposal \cite{Gisin} to 
use `local filters' to increase correlations between two entangled quantum 
subsystems. Following this a number of other schemes have been designed for 
the purpose of local purification \cite{Deutsch}. 
All of these have one idea in common:
they all rely on some form of classical communication on which
subsequent {\em post-selection} is based. This means that if we start
with an ensemble of $N$ pairs of particles in a mixed state, the final pure 
state will invariably have fewer particles. This will be seen 
as a consequence of the
fact that local operations (i.e. generalised filters) {\em cannot}
increase correlations. However, although the increase in correlations 
cannot be achieved, an error correction procedure can always be 
applied locally, which will maintain the entanglement. 

We introduce the necessary information--theoretic background in section 2.
In section 3 we present a simple model of atoms interacting `locally' with 
two entangled cavities and give a number of feedback
schemes by which the correlations might possibly be increased, 
without using any  classical communication and post-selection. 
We show that each of these
schemes fails, and we link this to the impossibility of superluminal
propagation of any signal. At the end of this section we briefly show how 
non--local interactions can easily be used to increase correlations. Section 4
presents two rigorous proofs of the impossibility of 
increasing correlations locally. In section 5 we derive general conditions
for error correcting codes, which are then used in section 6 to 
show that local error correction, in fact, preserves correlations
and entanglement. Using these considerations we then present a simple example
of how to encode two cavities against a single amplitude error on either cavity
using four atoms.   
  
\section{Theoretical Background}

In this section we introduce the information--theoretical background
necessary to understand the results in this paper. We summarise the 
basic definitions and mathematical framework relevant to the 
problem, and define the key concepts and quantities
which are used to characterise entanglement between systems (for a
more elaborate discussion of quantum information theory see references 
\cite{Ekert2,Everett}).

\subsection{Information Measures}

In this subsection we introduce various classical 
information measures \cite{Cover}.
Quantum analogues are then defined in the following subsection. 
Fundamental to our understanding of correlations is the measure 
of uncertainty in a given probability distribution.

\vskip 0.5cm 
{\em Definition} 1.
The uncertainty in a collection of possible states with corresponding 
probability distribution $p_i$ is given  by an {\em entropy}:
\begin{equation}
H(p) := -\sum_i p_i \ln p_i
\end{equation}
called {\em Shannon's entropy}. We note that there is no Boltzman constant
term in this expression, as there is for the physical entropy, since $k_B$
is by convention set to unity.
\vskip 0.5cm

We need a means of comparing two different probability distributions, and for 
this reason we introduce the notion of {\em relative entropy}.

\vskip 0.5cm 
{\em Definition} 2. 
Suppose that we have two probability distributions, ${p_i}$ and ${q_i}$.   
{\em The Shannon relative entropy} between these two distributions is defined as
\begin{equation}
D(p\,||\,q)  :=  -\sum_i p_i \ln \frac{p_i}{q_i}  \;  .
\end{equation}

This function is a good measure of the `distance' between ${p_i}$ and ${q_i}$, 
even though, strictly speaking, it is not a mathematical distance since 
$D(p\,||\,q)\ne D(q\,||\,p)$. Its information--theoretic significance becomes 
apparent through the notion of mutual information.

\vskip 0.5cm 
{\em Definition} 3.
{\em The Shannon mutual information} between two random variables $A$ and $B$, 
having a joint probability distribution ${p_{i,\alpha}}$, and
marginal probability distributions ${p_i}$ and ${p_{\alpha}}$ is defined 
as

\begin{equation}
I_S(A:B) := D(p_{i,\alpha}\,||\,p_i\, p_{\alpha})= H(p_i)+H(p_\alpha)-H(p_{i,\alpha})
\end{equation}
where the Latin indices refer to $A$ and Greek indices to $B$.
\vskip 0.5cm 

The Shannon mutual information, as its name indicates, measures the quantity 
of information conveyed about the random variable $A$ ($B$) through measurements 
of the random variable $B$ ($A$). Written in the above form, the Shannon mutual
information represents the `distance' between the joint distribution
and the product of the marginals; loosely speaking it determines
how far the joint state is away from the product state, and is hence 
suitable as a  measure of the degree of correlations between the two 
random variables. We now show how the above measure can be used to 
determine correlations between two `entangled' quantum systems.

\subsection{Quantum Correlations and Entanglement}

The general state of a quantum system is described by its density matrix $\hat{\rho}$.
If $\hat{A}$ is an operator pertaining to the system described by 
$\hat{\rho}$, then by the spectral decomposition theorem $\hat{A} =
\sum_i a_i \hat{P}_i$, where $\hat{P}_i$ is the projection onto the
state with the eigenvalue $a_i$. The probability of obtaining the
eigenvalue $a_j$ is given by $p_j = \mbox{Tr}(\hat{\rho} \hat{P}_j) = 
\mbox{Tr}(\hat{P}_j \hat{\rho})$. The uncertainty in a given observable
can now be expressed through the Shannon entropy. However, to 
determine the  uncertainty in the state as a whole we use the 
`von Neumann' entropy.

\vskip 0.5cm
{\em Definition} 4.
{\em The von Neumann entropy} of a quantum system described by a density 
matrix $\hat{\rho}$ is defined as
\begin{equation}
S(\hat{\rho}) := -Tr( \hat{\rho} \ln \hat{\rho} )
\end{equation}
\vskip 0.5cm
The Shannon entropy is equal to the von Neumann entropy only when it
describes the uncertainties in the values of a particular set of observables, 
called Schmidt observables \cite{Ekert2} (This is the set of observables 
that possesses the same spectrum as the density matrix describing the state).

\vskip 0.5cm
{\em Definition} 5.
The two quantum systems $A$ and $B$ are said to be {\it entangled} 
if their joint state cannot be expressed as a convex sum of the 
direct products of the individual states; otherwise they are disentangled (a
convex sum of direct products is a sum of the form $\sum_i p_i \hat\rho_A^i 
\otimes \hat\rho_B^i$, where indices $A$ and $B$ refer to the first and the 
second subsystem respectively, and $\sum_j p_j = 1$).
\vskip 0.5cm

The prime example of the entangled state is the EPR--type state between 
the two $2$--state systems, A and B:

\begin{equation}
\ket{\psi_{AB}} = \alpha\,\ket{0}_A\ket{1}_B+\beta\,\ket{1}_A\ket{0}_B \;  .
\label{epr}
\end{equation}
which obviously cannot be expressed as a direct product of the 
individual states, {\em unless} either $\alpha$ or $\beta$ equals
zero. 

To quantify the degree of correlations between the two quantum systems, 
we introduce the {\em von Neumann mutual information} via the notion of 
the reduced density matrix. If the joint state of the two quantum 
systems is $\hat{\rho}_{AB}$, then the reduced density matrices of the 
subsystems $A$ and $B$ are given by
\begin{eqnarray}
\hat{\rho}_A: =  \mbox{Tr}_B\,\hat{\rho}_{AB} \;\;\qquad
\hat{\rho}_B: =  \mbox{Tr}_A\,\hat{\rho}_{AB} 
\end{eqnarray}

In analogy with the Shannon relative entropy between two probability
distributions, we define the so called von Neumann relative entropy, as
a measure of `distance' between two density matrices.

\vskip 0.5cm
{\em Definition} 6.
Given two density matrices $\hat{\rho}_A$ and $\hat{\rho}_B$ the
{\em von Neumann relative entropy} is defined as: 
\begin{equation}
D(\hat{\rho}_A||\hat{\rho}_B) := -\mbox{Tr}\, \hat{\rho}_A \ln  
\frac{\hat{\rho}_A}{\hat{\rho}_B}
\end{equation}
(where $\ln \frac{\hat{\rho}_A}{\hat{\rho}_B} := \ln\hat{\rho}_A - \ln\hat{\rho}_B$).
  
The degree of correlation between the two quantum subsystems 
is given by the von Neumann mutual information, defined by analogy 
with the Shannon mutual information via the concept of relative entropy.

\vskip 0.5cm
{\em Definition} 7. 
{\em The von Neumann mutual information} between the two subsystems 
$\hat{\rho}_A$ and $\hat{\rho}_B$ of the joint state $\hat{\rho}_{AB}$ 
is defined as 
\begin{eqnarray}
I_N(\hat{\rho}_A:\hat{\rho}_B\, ;\hat{\rho}_{AB}) & := & D(\hat{\rho}_{AB}
\,||\,\hat{\rho}_A\otimes\hat{\rho}_B) 
\label{def7a}
\\ \nonumber \\
& = & S(\hat{\rho}_A) + S(\hat{\rho}_B) - S(\hat{\rho}_{AB})\;\;  .
\label{def7b}
\end{eqnarray} 
  
From this we can see that the state in eq. (\ref{epr}) is maximally 
correlated when $|\alpha|^2=|\beta|^2=\frac{1}{2}$, whereas the 
correlations are minimal for either $\alpha=0$ or $\beta=0$, i.e. 
when the state is disentangled. 

In this paper we mainly focus on two systems in a {\em joint pure 
state} \cite{Phoenix} in which case the entropy of the overall state, 
$S({\hat\rho}_{AB})$,  is zero, and the reduced entropies are equal \cite{Araki}. 
In this case, there are no classical uncertainties, and then the degree of 
correlation  is purely quantum mechanical. This is then also called the degree 
of entanglement. However, for mixed states it is at present not
possible to separate entirely quantum from classical correlations and a 
good measure of entanglement does not exist (although steps
towards resolution of this problem are being taken, e.g \cite{Horodecki}), 
which is the reason why we use the von Neumann mutual information throughout .    
  
\subsection{Entropic Properties}

In this subsection we present without proofs several properties of 
entropy which will be used in the later sections \cite{Wehrl1}. These are:
\begin{eqnarray}
&1.&\mbox{\em additivity:} \qquad \qquad \qquad \,\,\, S(\hat{\rho}_A\otimes
\hat{\rho}_B) = S(\hat{\rho}_A) + S(\hat{\rho}_B) ;
\label{ep1}
\\
&2.&\mbox{\em concavity:} \qquad \qquad \qquad \,\,\, S \left ( \sum_i 
\lambda_i \hat{\rho}_i \right ) \ge \sum_i \lambda_i S(\hat{\rho}_i);
\label{ep2}
\\
&3.&\mbox{\em strong subadditivity:} \qquad S(\hat{\rho}_{ABC}) + S(\hat{\rho}_B) 
\le S(\hat{\rho}_{AB}) + S(\hat{\rho}_{BC}).
\label{ep3} 
\end{eqnarray}
(where $\hat{\rho}_B = \mbox{Tr}_{AC} \hat{\rho}_{ABC}$ and similarly for the others).

It is also worth mentioning that the consequence of the strong subadditivity 
is the so called weak subadditivity described by the Araki--Lieb inequality 
\cite{Araki}: $S(\hat{\rho}_{AB}) \le S(\hat{\rho}_A) + S(\hat{\rho}_B)$ . 
This asserts that there is less uncertainty in the joint state 
of any two subsystems than if the two subsystems are considered separately. 
We now turn to describing two equivalent ways of {\em complete measurement}.

\subsection{Complete Measurement}

In this subsection we present two different ways to describe the 
dynamical evolution of a quantum system. First we can look at the joint
unitary evolution of the system, $S$, and its environment, $E$. The environment
can be a similar quantum system to the one we observe, or much larger: we
leave this choice completely open in order to be as general as possible.
Let the joint `$S$+$E$' state initially be disentangled, $\ket{\psi}_S
\ket{\psi}_E$, after which we apply a unitary evolution $\hat U_{SE}$
on `$S$+$E$' resulting in the state

\begin{equation}
\hat U_{SE}\ket{\psi}_S\ket{\psi}_E
\label{uni}
\end{equation}

Since we are interested in the system's evolution only, to obtain its
final state, $\hat\rho_S$, we have to trace over the environment, i.e.
\begin{equation}
\hat{\rho}_S = \mbox{Tr}_E ( \hat{U}_{SE}\,\ket{\psi}_S\bra{\psi}_S\otimes 
\ket{\psi}_E\bra{\psi}_E \, \hat{U}_{SE}^{\dagger})
\end{equation}

Another way to obtain the same result is to exclude the environment from the
picture completely by defining operators of the `complete measurement' \cite{Davies}
\begin{equation}
\sum_{i} \hat{A}^{i\dagger} \hat{A}^i = \hat{{\it I}}
\label{comp}
\end{equation}

which act on the system alone, and therefore to be equivalent to
the above system's evolution must satisfy
\begin{equation}
\sum_{i}  \hat{A}^i \ket{\psi}_S\bra{\psi}_S\hat{A}^{i\dagger}= 
\hat{\rho}_S \; .
\end{equation}
Let us now derive the necessary form of $\hat A$'s using eq. (\ref{uni}). 
Let an orthonormal basis of $E$ be $\{\ket{\phi}^i_E\}$. Then,
\begin{equation}
\hat{A}^i = \bra{\phi}^i_E \, \hat{U}_{SE} \ket{\psi}_E
\label{cm1}
\end{equation}
It can easily be checked that the above $\{\hat A^j\}$'s satisfy the 
completeness relations in eq. (\ref{comp}). Since the choice of 
basis for $E$ is not unique, then neither is the choice of complete 
measurement operators. In fact, there is an infinite number of 
possibilities for the operators $\{ \hat A^j\}$. Note that the dimension 
of the complete measurement, $\hat A$, is in general different to the 
dimension of the observed system, and in fact equal to the dimension 
of $E$.

\section{Atom--Cavity Models}

We present here a simple model which aims to {\it increase} the quantum 
correlations between two entangled subsystems.  The model we present
employs a technique of performing `local' complete measurements. By 
this, we mean that when the two quantum systems are entangled we 
perform complete measurements on either subsystem separately, while 
not interacting directly with the other subsystem.  We may regard this
result to be counter-intuitive --- it does not seem at first sight
possible that purely local operations could increase the non-local 
quantum features. There  have been many schemes devised whereby 
correlations can be increased by local measurements on an ensemble 
of systems combined with {\it classical communication}, followed by 
a procedure of {\it post-selection}. Indeed, the model presented here 
can also be adapted readily to represent such 
a scheme.  However, we verify that by local measurement alone, 
and without post-selection based upon classical communication, the 
correlations do not increase.  In the next section we present two
proofs that this is, in fact, a general result.

The models used to demonstrate this are of the `cavity QED' type, 
and are both easy to understand physically and simple to analyse 
analytically. A good outline
of cavity QED is given in \cite{haroche}. We consider two optical cavities, 
the field states of which are entangled number states (for simplicity)
\begin{eqnarray}
\ket{\Psi}_{AB} = \alpha\ket{n}_A\ket{m}_B + \beta\ket{n^{\prime}}_A\ket{m^{\prime}}_B,
\label{m1}
\end{eqnarray}
where the subscripts `A' and `B' refer to the two cavities, and without loss 
of generality, we assume that $|\alpha |$ $>$ $|\beta |$. This is a pure 
state but is not maximally entangled.  The aim is to produce the state:
\begin{eqnarray}
\ket{\Phi}_{AB} = \frac{1}{\sqrt{2}}\left ( \ket{n}_A\ket{m}_B + 
\ket{n^{\prime}}_A\ket{m^{\prime}}_B \right ),
\label{m2}
\end{eqnarray}
i.e. we have made $\alpha$=$\beta$=$\frac{1}{\sqrt{2}}$, which is maximally entangled.

Two-level atoms are sent, one at a time, through cavity $A$ and interact with that individual 
cavity field, via the Jaynes-Cummings Hamiltonian \cite{Shore}, for a pre-determined time 
period. After each atom passes through the cavity, a measurement is made which 
projects the atomic state into either the ground state or the excited state.  
Due to the entanglement developed between the atom and the field in cavity $A$ 
during the interaction, this measurement also collapses the joint cavity $A$ -- cavity 
$B$ field state into a different superposition, 
one with either the same number of photons in cavity $A$, or with one 
extra photon respectively. By successively sending atoms through the 
cavity for interaction periods determined from the state of the previously 
measured atom, a {\it feedback} mechanism can be set up whereby one might 
expect to optimise the probability of achieving the state defined in eq.(\ref{m2}).
Similar schemes have been used on single cavities for quantum state-engineering 
\cite{Garraway}.

We also consider extensions to this procedure. Firstly, we mention procedures for 
interacting locally with both cavities, the qualitative results of which are the same.
And secondly, we give two examples of non-local interactions, which give quite different 
results to the above local procedures.

\subsection{Cavity Models With Local Feedback}

The first model involves sending atoms through cavity $A$ only, a schematic
of which is given in Fig.1; we assume the initial joint cavity field state 
is given by (\ref{m1}).  The first
atom is in the excited state, and so the initial atom-field state is
\begin{eqnarray}
\left ( \alpha\ket{n}_A\ket{m}_B + \beta\ket{n^\prime}_A\ket{m^\prime}_B 
\right )\otimes\ket{e}_A.
\label{m3}
\end{eqnarray}
After interaction for a time $t_1$, determined from the atomic time of flight, 
the joint atom-field state becomes
\begin{eqnarray}
\left ( \alpha a_n(t_1)\ket{n}_A\ket{m}_B + 
        \beta a_{n^\prime}(t_1)\ket{n^\prime}_A\ket{m^\prime}_B \right ) 
       \otimes \ket{e}_A
\nonumber
\\
\nonumber
\\
+
\left ( \alpha b_n(t_1)\ket{n+1}_A\ket{m}_B + 
        \beta b_{n^\prime}(t_1)\ket{n^\prime+1}_A\ket{m^\prime}_B \right ) 
       \otimes \ket{g}_A
\label{m4}
\\
\nonumber
\end{eqnarray}
where the coefficients are given by
$a_n(t_1) = \cos \left ( \frac{R_nt_1}{2}\right )$, 
$a_{n^\prime}(t_1)=\cos \left ( \frac{R_{n^\prime}t_1}{2}\right )$, 
$b_n(t_1) = -i \sin \left ( \frac{R_nt_1}{2}\right )$,
$b_{n^\prime}(t_1) = -i \sin \left ( \frac{R_{n^\prime}t_1}{2}\right )$,
and $R_i=2g\sqrt{i+1}  =  R_o\sqrt{i+1}$.

We now arrange that the velocity of the atom, and hence the 
interaction time with the field, is such that
\begin{eqnarray}
\alpha a_n(t_1) = \beta a_{n^\prime}(t_1),
\label{m8}
\end{eqnarray}
in which case the joint atom-field state becomes
\begin{eqnarray}
\alpha a_n(t_1) \left ( \ket{n}_A\ket{m}_B + 
\ket{n^\prime}_A\ket{m^\prime}_B \right ) 
\otimes\ket{e}_A
\nonumber
\\
\nonumber
\\
+
\left ( \alpha b_n(t_1) \ket{n+1}_A\ket{m}_B + 
\beta b_{n^\prime}(t_1)\ket{n^\prime+1}_A\ket{m^\prime}_B \right ) 
\otimes \ket{g}_A
\label{m9}
\end{eqnarray}
From this we see that if we measure the atom in the excited state, the
resulting cavity field state is maximally entangled.  The probability
of measuring the excited atomic state is
\begin{eqnarray}
P_1(e)=2|\alpha a_n(t_1)|^2.
\label{m10}
\end{eqnarray}
If we were to prepare a whole ensemble of cavities in precisely the same initial
state (\ref{m3}), then after measurement on all of the ensemble members, we
would have prepared approximately $(100\times P_1(e))\%$ of the cavities in the maximally
entangled state (\ref{m2}).  We can discard all the cavities for which we measured
the atom in the ground state, and we will have a whole sub-ensemble of cavities for which
the entanglement has increased.  This is the post-selection procedure mentioned
earlier, and always requires that measurements on the whole ensemble be `thrown away' 
in order to increase the entanglement of a sub-ensemble.

What we wish to do here is to increase the entanglement on an {\it individual pair} 
of entangled cavities. Instead of performing one measurement on an ensemble of cavities, 
we keep performing a number of measurements on this single pair until we achieve our aim.  
When the atom is measured in the excited state, we are there. If the outcome of the atomic 
state measurement was $\ket{g}_A$, the final cavity field state would be the 
corresponding field state in eq.(\ref{m9}), which is still entangled, but not 
maximally so. We can now use this field state as 
a {\it new} initial entangled cavity field. In this way, we would hope that it is 
just a matter of sending through `enough' atoms until the desired state is reached.  

Since the field state corresponding to a ground state measurement involves 
the $(n+1)$ Fock state, sending through another excited atom allows the possibility 
of generating an $(n+2)$ Fock state, which takes us further away from the initial 
state (\ref{m3}).  We thus send through a ground-state atom, which can remove 
the extra photon.  

Using, therefore, this as the starting field-state, we define the `new'
$\alpha$ and $\beta$ as
\begin{eqnarray}
{\alpha}^\prime =\frac{ \alpha b_n(t_1)}{\sqrt{({\alpha b_n(t_1)})^2+
({\beta b_{n^\prime}(t_1)})^2}}, \qquad 
{\beta}^\prime = \frac{ \beta b_{n^\prime}(t_1)}{\sqrt{({\alpha b_n(t_1)})^2+
({\beta b_{n^\prime}(t_1)})^2}}.
\label{m13}
\end{eqnarray}
and the joint atom-field state after sending through a ground state atom 
for time $t_2$, such that $b_{n}(t_2) {\alpha}^\prime = b_{n^\prime}(t_2){\beta}^\prime$, 
becomes
\begin{eqnarray}
\left (  a_{n}(t_2) {\alpha}^\prime\ket{n+1}_A\ket{m}_B  
+ a_{n^\prime}(t_2){\beta}^\prime\ket{n^\prime+1}_A\ket{m^\prime}_B \right ) 
\otimes \ket{g}_A
\nonumber
\\
\nonumber
\\
+\,
b_{n}(t_2) {\alpha}^\prime 
\left ( \ket{n}_A\ket{m}_B + \ket{n^\prime}_A\ket{m^\prime}_B \right ) 
\otimes \ket{e}_A
\label{m16}
\end{eqnarray}
As before, if the atom is measured in the excited state, then the cavities
are left in the maximally entangled field state, once normalised, as desired.
The probability for this measurement is
\begin{eqnarray}
P_2(e)=2| b_{n}(t_2) {\alpha}^\prime |^2.
\label{m17}
\end{eqnarray}
It is worth noting at this point that the state of the field, after measuring 
a ground state atom, is in itself {\it less} entangled than the initial state 
(\ref{m1}). This is a direct consequence of the concave property of entropy 
when applied to either reduced density matrix. Namely, the fact that in one 
case, when registering an excited atom, the field becomes more entangled than 
previously (i.e. the entropy of either reduced system is greater after the 
interaction), implies that the entanglement of the other field state, when 
we register a  ground atom, is `smaller' than previously (i.e. the entropy 
is smaller than  before the interaction). This can be quantified as follows.  
Let the reduced field state after the interaction be
\begin{eqnarray}
{\hat\rho}_A^\prime = \mbox{p}{\hat\rho}_{A1}^\prime + \mbox{(1 - p)}{\hat\rho}_{A2}^\prime
\label{m17.1}
\end{eqnarray}
where ${\hat\rho}_{A}^\prime$ is the reduced density matrix for cavity $A$ formed
from eq.(\ref{m9}), and ${\hat\rho}_{A1}^\prime$, ${\hat\rho}_{A2}^\prime$ are the
parts of ${\hat\rho}_{A}^\prime$ corresponding to the measurement of an excited
or ground state atom respectively.  Now using the concave property eq.(\ref{ep2})
we see that
\begin{eqnarray}
S({\hat\rho}_{A}) = S({\hat\rho}_{A}^\prime) \ge \mbox{p} S({\hat\rho}_{A1}^\prime) +
                                         \mbox{(1 - p)} S({\hat\rho}_{A2}^\prime)
\label{m17.2}
\end{eqnarray}
where the first equality follows from the fact that the reduced density
matrix does not change during this interaction, which can readily be derived 
for this example, and is shown generally in the next section.  It follows that
\begin{eqnarray}
S({\hat\rho}_{A}) \ge \mbox{p} (S({\hat\rho}_{A}) + \Delta) + 
                  \mbox{(1 - p)} S({\hat\rho}_{A2}^\prime)
\label{m17.3}
\end{eqnarray}
where $\Delta$ is the amount by which the entropy (and hence entanglement) of the 
reduced subsystem is constructed to increase upon measurement of $\ket{e}$, by 
arranging atomic interaction times. So,

\begin{eqnarray}
S({\hat\rho}_{A}) - S({\hat\rho}_{A2}^\prime) \ge \frac{\mbox{p}\Delta}{\mbox{1 - p}} > 0
\label{m17.4}
\end{eqnarray}
Hence, it is immediately seen that
\begin{eqnarray}
S({\hat\rho}_{A}) > S({\hat\rho}_{A2}^\prime)
\label{m17.5}
\end{eqnarray}
and the result (in this case) is proven.  

A small amount of simple algebra applied to eq.(\ref{m8}) shows that whatever the 
initial values  of $\alpha$ and $\beta$, the ratio
\begin{eqnarray}
\frac{\min(\alpha,\beta)}{\max(\alpha,\beta)}
\label{m12}
\end{eqnarray}
always {\it decreases} unless $n=n^\prime$ i.e. the cavities are not entangled
in the first place (a ratio equal to unity implies maximal entanglement).  We 
thus have that $|{\alpha}^\prime | > |\alpha |$ and $|{\beta}^\prime | < |\beta |$.  
It is readily seen from this, and the fact that $|a_i(t)|<1$ and $|b_i(t)|<1$, that
\begin{eqnarray}
P_2(e)_{\mbox{max}}=2|{\beta}^\prime|^2 \qquad < \qquad P_1(e)_{\mbox{max}}=2|{\beta}|^2
\label{m18}
\end{eqnarray}
Thus, there are two effects each time an atom is sent through the cavity --- the  
first is that the probability of detecting an atom in the excited state, and 
hence collapsing the field state to the maximally entangled form, on average 
decreases with each atom that goes through; and the second is that the field-state 
if the atom is measured in the ground state becomes successively more disentangled.
The effect is to make it successively more likely that the field will become 
completely disentangled, rather than completely entangled, which was the original aim.
This can be seen mathematically by adding up the probabilities of detecting an atom
in the excited state after sending through exactly $N$ atoms.  If the probability
of detection in state $\ket{e}$ after the $i$-th atom is $a_i$, and the 
corresponding probability for $\ket{g}$ is $b_i$, then the probability of
detection in $\ket{e}$ after $N$-atoms is
\begin{eqnarray}
&&a_0 + b_0a_1 + b_0b_1a_2 + b_0b_1b_2a_3 + b_0b_1b_2b_3a_4 + ... + b_0...b_{N-1}a_N
\nonumber
\\
&=& (1-b_0) + b_0(1-b_1) + b_0b_1(1-b_2) + b_0b_1b_2(1-b_3) + ... + b_0...b_{N-1}(1-b_N)
\nonumber
\\
&=& 1 - \prod_{i=0}^{N} b_i
\label{m18.1}
\end{eqnarray}
The above product term is always less than unity since each and every $b_i$ is 
individually less than unity, and similarly is always positive since all $b_i$ are 
individually positive, so the probability of detection of $\ket{e}$ after 
$N$-atoms is less than unity. In the limit of $N\rightarrow \infty$, it can be 
verified by a computer program that the above product always tends to the value 
of $2|\beta|^2$.  This result has the following consequence. 
In the limit $N\rightarrow \infty$ we either register a maximally entangled state or
a completely disentangled state. However we could arrange the atom cavity
interaction time to be such that this happens when the first atom goes
through the cavity. In this case it can be easily shown that the
probability for the maximally entangled state to be registered (i.e. measuring the
excited atomic state) is exactly $2|\beta|^2$.  Thus, no matter how many atoms 
we send through the cavity (one or infinitely many), the highest probability of 
reaching of reaching the maximally entangled state is {\it always} less than unity.  
We thus see that this scheme cannot increase correlations between two entangled systems.

We note also that we do not have to aim to achieve maximum entanglement for the 
particular initial state given by eq.(\ref{m3}).  We could continue to send,
for example, excited atoms and simply hope to achieve increased entanglement for
{\it any} state.  However, the same arguments given above also show that we 
cannot increase the entanglement of {\it both} field states corresponding to
the two atomic measurement outcomes, as eq.(\ref{m17.5}) shows. 

We should note that if it was possible to increase entanglement by the above 
local scheme, we would have a means of superluminal communication. Namely, the 
sender of the message could change the entanglement by operating locally on his 
cavity which could then be detected on the other end by the receiver in possession 
of the other cavity. The communication would then proceed as follows: two 
participants would initially share a number of not maximally entangled cavities. 
Then, if the sender does nothing on one of his cavities, this could represent 
`logical zero', whereas if the sender maximally entangled the cavities this would 
represent `logical one'. After sharing the entangled sets of cavities, the two 
participants could travel spatially as far away from each other as desired.  In 
this way, they would be able to communicate, through the above binary code, at 
a speed effectively instantaneously (only the time to actually prepare the binary 
states, and to measure them at the other end).  Therefore, we see that the 
impossibility of locally increasing the correlations is closely related to 
Einstein's principle of causality. This is a curious consequence of quantum 
mechanics, the postulates of which contain no reference to special relativity. 
Indeed, this could be turned upside down, and viewed as one reason why the 
above (or any similar) scheme would not work.

We thus find that the above scheme cannot increase correlations by local
actions on one cavity alone. We might expect to compensate for this by sending 
independent atoms through both cavities, and arranging a feedback mechanism based 
upon classically communicating the knowledge of each state to the other side. In 
this way, we approach more closely the scheme of classical communication 
with post selection \cite{Gisin}, but hope to replace the post-selection procedure with
that of sending through multiple atoms until we achieve success.  We would also 
expect to avoid superluminal communications since the method inherently involves 
classical communication between the two observers. The analysis for this problem 
is very similar to that given above for the one-atom model, except that there 
is much more freedom to choose which state to measure and how to optimise it. 
Following through a similar reasoning as in the single atom model, it
is readily deduced that there is no way in this scheme to increase
correlations. There are numerous variations on this above scheme:
maximising the probability of detection in $\ket{e}_A\ket{e}_B$, 
minimising the rate of change of $\alpha$ and $\beta$, and so on, but the basic 
fact that the probability is never identically unity for any number of atoms 
remains the same.

\subsection{Increasing Entanglement Non-Locally}

We now present two simple examples showing how a {\it nonlocal} operation 
{\it can} increase and, in fact, create correlations and entanglement.  The 
procedures described here can be used to prepare initially entangled states.

\subsubsection{Method 1}
Suppose that the two cavities, A and B, start disentangled in the state:
\begin{equation}
\ket{{\phi}_{cav}}_{AB} = \frac{1}{\sqrt{2}} \ket{0}_A \ket{0}_B  \; .
\end{equation}
Let us send an entangled atomic pair through the cavities, each atom
going through one cavity only, with the initial atomic state:
\begin{equation}
\ket{{\phi}_{atom}}_{AB} = \frac{1}{\sqrt{2}} ( \ket{e}_A \ket{g}_B + \ket{g}_A \ket{e}_B)\; .
\end{equation}
After the interaction for the same time $t$ the joint state will be:
\begin{eqnarray}
\ket{{\psi}_{joint}}_{AB} & = & 
-b^2_o(t) \, \ket{g}_A \, \ket{g}_B 
\left \{ 
\frac{1}{\sqrt{2}} ( \ket{0}_A \ket{1}_B + \ket{1}_A \ket{0}_B) 
\right \} 
\nonumber
\\
& + & 
a^2_o(t) \, 
\left \{
\frac{1}{\sqrt{2}}(\ket{e}_A \ket{g}_B + \ket{g}_A \ket{e}_B )
\right \}
\, \ket{0}_A\,\ket{0}_B \; .
\end{eqnarray}
Therefore, by simply setting $ b_o(t) = 0$ we end up with 
certainty in the maximally entangled field state. Hence nonlocal
interactions can, as expected, increase and create correlations and
entanglement.  The difference between this scheme and the previous two
is that entanglement is being transferred to the cavities, from the
atoms.  This allows the cavity entanglement to `increase', but at the
expense of the entanglement of the atoms.

\subsubsection{Method 2}

This method involves only one atom, first interacting with one cavity and 
then with the other. This type of ``entanglement generation" has been 
analysed in a number of other places \cite{Meystre}. Let the initial state of 
`atom+fields' be:
\begin{equation}
\ket{e} \ket{0}_A \ket{0}_B
\end{equation}
After interaction between the atom and the cavity $A$ for time $t_1$ the state is
\begin{equation}
(a_o(t_1) \ket{e}\ket{0}_A + b_o(t_1)\ket{g}\ket{1}) \ket{0}_B \;\;   .
\end{equation}
The atom now interacts with the cavity $B$ for time $t_2$ after which the final state is
\begin{equation}
a_o(t_1) a_o(t_2) \ket{e}\ket{0}_A\ket{0}_B + a_o(t_1) b_o(t_2) \ket{g}\ket{0}_A
\ket{1}_B + b_o(t_1) \ket{g}\ket{1}_A\ket{0}_B  \;\; .
\end{equation}
Choosing $a_o(t_2) = 0$ and $a_o(t_1) = b_o(t_2) = \frac{1}{\sqrt{2}}$ the
above reduces to:
\begin{equation}
 \ket{g}\frac{1}{\sqrt{2}}(\ket{0}_A\ket{1}_B + \ket{1}_A\ket{0}_B)\;\; ,
\end{equation}
which is the desired, maximally entangled state of the field.  Thus, the
method achieves an entangled cavity state by creating an entangled 
atom-cavity state, and transferring this to the two cavities alone.

\section{Local Interactions Cannot Increase Correlations}

The central problem addressed in this paper, and described in the
specific examples in the previous section, is summarised
in the following theorem:
\vskip 0.5cm

{\bf Theorem}. Correlations do not increase during local 
complete measurements carried on two entangled quantum systems.
\vskip 0.5cm

We present here two quite separate, but mathematically rigorous 
proofs of this theorem, the first using the notion of entropy,
the second using the ideas of complete measurements and conditional 
entropy as a measure of relative information.  

First, we show that, as mentioned in subsection 3.1, no local complete 
measurement on subsystem $A$ can change the reduced density matrix 
of $B$, and vice versa. Let us perform a complete measurement on $A$, 
defined by $\sum_i \hat A^{i\dagger}\hat A^i\otimes \hat I_B = \hat I$ where 
the identity in the direct product signifies that the other subsystem does 
not undergo any interaction. Let the overall state of `$A$+$B$' be described 
by $\hat\rho$. Then after $A$ has undergone a complete measurement, $B$'s 
reduced density matrix is given by:
\begin{eqnarray}
\hat\rho_B^\prime & = & tr_A \left \{ \sum_i \hat A^i\otimes \hat I_B \,\,\hat\rho 
\,\, \hat A^{i\dagger}\otimes \hat I_B \right \} \nonumber \\
        &=&  \sum_i tr_A \{\hat \rho \hat A^{i\dagger} \hat A^i\otimes \hat I_B \} 
        =  tr_A \left \{\hat \rho \sum_i \hat A^{i\dagger} \hat A^i \otimes 
           \hat I_B \right \} = tr_A \{\hat \rho\}
        = \hat{\rho}_B \,\, .
\end{eqnarray}
Therefore the equality in eq. (\ref{m17.2}) is now justified.  We note also that
the $\hat A$'s in the above equation can be unitary operators, since 
$\hat U^{\dagger}\hat U=\hat I$. We use this result in {\em Proof.1} below.

\subsection{{\it Proof 1:}}
This proof is due to Partovi \cite{Partovi89}, who proved it as a general 
result, rather than applied it to increasing correlations by local operations.
Consider three quantum systems $A$, $B$, $C$, initially in the state
described by a density matrix of the form: $\hat \rho_{ABC}(0)=\hat \rho_{AB}(0)
\hat \rho_{C}(0)$, i.e. A and B are initially correlated and both completely 
independent from $C$. We are now going to let $B$ and $C$ interact and 
evolve unitarily for time $t$, resulting in the state $\hat \rho_{ABC}(t)$. 
The partial trace is defined in the usual fashion, e.g. $\hat \rho_{AB}(t) = 
\mbox{tr}_C\hat \rho_{ABC}(t)$, and similarly for all the other subsystems.  
Now we use the strong subadditivity \cite{Partovi89} applied to $A+B+C$ 
at time $t$ to obtain

\begin{equation}
S_{ABC}(t) + S_{B}(t) \le S_{AB}(t) + S_{BC}(t) \;\;  .
\label{3.1}
\end{equation}

But $S_{ABC}(t)=S_{ABC}(0)$, as the whole system evolves unitarily.
Also, $S_{ABC}(0)=S_{AB}(0)+S_{C}(0)$, since at the beginning $C$ is 
independent from $A, B$. $A$ is only a spectator in the evolution of $B$ and 
$C$, so that, as shown above, $S_{A}(t)=S_{A}(0)$, $S_{BC}(t)=S_{BC}(0)$. Finally, 
there are no correlations between $B$ and $C$ at the beginning, implying: 
$S_{BC}(0)=S_{B}(0)+S_{C}(0)$. Invoking the definition in eq.(\ref{def7b}) 
for quantum correlations, and using the above properties and strong 
subadditivity in eq.(\ref{3.1}), we arrive at the following

\begin{equation}
I({\hat\rho}_A:{\hat\rho}_B;{\hat\rho}_{AB})(t) \le I({\hat\rho}_A:{\hat\rho}_B;
{\hat\rho}_{AB})(0)
\label{3.2}
\end{equation}   

Adding another system $D$ to interact with $A$ locally would lead to the 
same conclusion, hence completing the proof.

\subsection{{\it Proof 2:}}
This proof is a quantum analogue of the well known classical result that can 
loosely be stated as `Markovian processes cannot increase correlations' 
\cite{Everett,Cover}. 
We will now describe the interactions of $A+B$ with $C$ and $D$ in terms
of complete measurement's performed on $A+B$. Let the state 
of $A+B$ be initially described by the density operator $\hat \rho$, whose 
diagonal elements, $\rho_{ii}$, give the probabilities of being in various 
states, depending on the basis of the density matrix. Let this state undergo 
a complete measurement, described by operators $\hat A^j$, such that 

\begin{equation}
\sum \hat A^{\dagger\, j} \hat A^j = {\hat I}    \; .
\label{3}
\end{equation}

The new diagonal elements are then:

\begin{equation}
\rho_{ii}^{\prime} = \sum_{nlm} A^{n}_{il}\rho_{lm} {A^{\dagger}}^n_{mi} \; .
\label{4}
\end{equation}

Let us introduce a relative information measure (defined in sec.2) to 
$\rho_{ii}$: to each value of $\rho_{ii}$ we assign a nonnegative number 
$a_{ii}$. We now wish to compare the distance \cite{Everett} between 
$\hat \rho$ and $\hat a$ before 
and after ($\hat {\rho}^{\prime}$ and $\hat a^{\prime}$) the complete 
measurement, $\hat A$. The distance after the measurement is:

\begin{eqnarray}
\sum_i \rho_{ii}^{\prime} \log \frac{\rho_{ii}^{\prime}}{a^{\prime}_{ii}} 
& = & \sum_i  \left (\sum_{nlm} A^{n}_{il}\,\rho_{lm}\, {A^{\dagger}}^n_{mi} \right ) 
\,\log \frac{\sum_{nlm} A^{n}_{il}\,\rho_{lm}\, 
{A^{\dagger}}^n_{mi}}{\sum_{nlm} A^{n}_{il}\, a_{lm}\, {A^{\dagger}}^n_{mi}} \nonumber \\
& \le & \sum_i  \left (\sum_{nlm} A^{n}_{il}\,\rho_{lm}\, {A^{\dagger}}^n_{mi} \right ) 
\,\log \frac{A^{n}_{il}\,\rho_{lm}\, 
{A^{\dagger}}^n_{mi}}{A^{n}_{il}\, a_{lm}\, {A^{\dagger}}^n_{mi}}\nonumber \\
& = &   \sum_i  \left (\sum_{nlm} A^{n}_{il}\,\rho_{lm}\, {A^{\dagger}}^n_{mi} \right ) 
\,\log \frac{\rho_{lm}}{a_{lm}} \nonumber \\
& = & \sum_{lm} \rho_{lm} \left (\sum_{in} A^n_{il} \, {A^{\dagger}}^n_{mi} \right )\, 
\log \frac{\rho_{lm}}{a_{lm}} \nonumber \\
& = & \sum_{lm} \rho_{lm} \delta_{lm} \log \frac{\rho_{lm}}{a_{lm}} \nonumber \\
& = & \sum_l \rho_{ll} \log \frac{\rho_{ll}}{a_{ll}}  \,\,  ,
\label{5}
\end{eqnarray} 
where for the inequality in the second line we have used one of the 
consequences of the concave property of the logarithmic function \cite{Everett,Wehrl1}, and in the fifth line we used the  
completeness relation given in eq. 
(\ref{3}). This implies that the distance  between the density matrix 
distribution and the relative information measure  decreases by making 
a complete measurement. If we now consider the particular case where 
${\hat a}$ is taken to be a distribution generated by the direct  product 
of the reduced density matrices (i.e. if we assume no correlations),  then 
the result above implies that the full density matrix becomes `more like' 
the uncorrelated density matrix.  From this, the theorem immediately follows.

\section{Conditions For Quantum Error--Correcting Codes}

We now describe an alternative way of manipulating quantum states
which can be described using the language of quantum computation \cite{Barenco}.
Quantum computation involves unitary operations and measurement 
on `quantum bits', or {\it qubits}.  A classical bit represents one 
of two distinguishable states, and is denoted by a `0' or a `1'.  
On the other hand, a qubit is in general a {\it superposition} of 
the two states, $\alpha\ket{0}+\beta\ket{1}$, and its evolution is
governed by the laws of quantum mechanics, for which a closed system evolves
unitarily.  This is reflected in the nature of the elementary gates, which
must be reversible, i.e. a knowledge of the output allows inference of the
input.  The practical realisation of a qubit can be constructed from
any two-state quantum system e.g. a two-level atom, where the unitary 
transformations are implemented through interaction with a laser.
An advantage of quantum computation lies in the fact that the input
can be in a coherent superposition of qubit states, which are then simultaneously
processed. The computation is completed by making a measurement on the output.  
However, a major problem is that the coherent superpositions 
must be maintained throughout the computation.
In reality, the main source of coherence loss is due to dissipative coupling to an
environment with a large number of degrees of freedom, which must be traced 
out of the problem.  This loss is often manifested as some form of
spontaneous decay, whereby quanta are randomly lost from the system.
Each interaction with, and hence dissipation to, the environment can be
viewed in information theoretic terms as introducing an error in the 
measurement of the output state.  There are, however, techniques for
`correcting' errors in quantum states \cite{Steane1,Steane2,Calderbank1}.
The basic idea of error-correction is to introduce an excess of
information, which can then be used to recover the original state
after an error. These quantum error correction procedures are in 
themselves quantum computations, and as such also susceptible to the 
same errors.  This imposes limits on the nature of the `correction codes',
which are explored in this section.  In the context of the present
paper, we will use error-correction as a way of maintaining coherence
between entangled cavities, described in the next section.  

First we derive general conditions which a quantum error correction 
code has to satisfy and are, in particular, less restricting than 
those previously derived in \cite{Ekert1}. Our derivation is an alternative to that in \cite{Knill1}, which also arrives at the same conditions. 
We use the notation of Ekert and Macchiavello\cite{Ekert1}. Assume that 
$q$ qubits are encoded in terms of $n$ qubits to protect against a certain 
number of errors, $d$. We construct $2^q$ {\em code--words}, 
each being a superposition of states having $n$ qubits. These code-words
must satisfy certain conditions, which are derived in this section. There 
are three basic errors \cite{Ekert1} (i.e. all other errors can be written 
as a combination of those): amplitude, $\hat A$, which acts as a NOT gate; phase, 
$\hat P$, which introduces a minus sign to the upper state; and their
combination, $\hat A\hat P$. A subscript shall designate the position of 
the error, so that $\hat P_{1001}$ means that the first and the fourth
qubit undergo a phase error. 

We consider an error to arise due to the interaction of the system with
a `reservoir' ({\it any} other quantum system), which then become entangled. 
This procedure is the most general 
way of representing errors, which are not restricted to discontinuous
`jump' processes, but encompass the most general type of interaction. 
Error correction is thus seen as a process of disentangling the system 
from its environment back to its original state. 
The operators $\hat A$ and $\hat P$ are constructed to operate only on the 
system, and are defined in the same way as the operators for a complete 
measurement described in subsection 2.4, eq.(\ref{cm1}).  In reality,
each qubit would couple independently to its own environment, so the
error on a given state could be written as a direct product of the
errors on the individual qubits.  A convenient error basis for a single
error on a single qubit is $\{ \hat I, \hat {\sigma}_i \}$, where the
$\hat {\sigma}_i$'s are the Pauli matrices.  In this case,
the error operators are Hermitian, and square to the identity operator,
and we assume this property for convenience throughout the following 
analysis. 

In general the initial state can be expressed as
\begin{equation}
	|\psi_i\rangle = \sum_{k=1}^{2q} c_k|C^k\rangle \ket{R}
	\label{6}
\end{equation}
where the $|C^k\rangle$ are the code--words for the states $|k\rangle$ and
$\ket{R}$ is the initial state of the environment. 
The state after a general error is then a superposition of all possible errors acting 
on the above initial state
\begin{equation}
	|\psi_f\rangle = \sum_{\alpha\beta} \hat A_{\alpha}\hat P_{\beta} \sum_{k}
	c_k|C^k\rangle\ket{R_{\alpha,\beta}} .
	\label{7a} 
\end{equation}
where $\ket{R_{\alpha,\beta}}$ is the state of the environment. Note that 
$\ket{R_{\alpha,\beta}}$ depends only on the nature of the errors, and is
{\em independent} of the code--words \cite{Ekert1}. The above is, in
general, not in the Schmidt form, i.e. the code--word states after the 
error are not necessarily orthogonal (to be shown) and neither are 
the states of the environment. Now, since we have no information 
about the environment, we must trace it out using an orthogonal basis for the 
environment $\{\ket{R_n},n=1,d\}$. The resulting state is a mixture
of the form $ \hat\eta_i = \sum_n \ket{\psi_n}\bra{\psi_n}$, where

\begin{equation}
\ket{\psi_n} = \sum_{\alpha\beta} x_n^{\alpha\beta} \hat{A}_{\alpha} \hat{P}_{\beta} \, \sum_k c_k \ket{C^k} \,\,\, ,
\label{7b}
\end{equation}
and $ x_n^{\alpha\beta} = \bra{R_n}R_{\alpha\beta}\rangle$.  To detect an error, one then 
performs a measurement on the state $\hat\eta$ to determine whether it has an overlap with 
one of the following subspaces 
\begin{equation}
	{\cal H}_{\alpha\beta} = \{ \hat A_{\alpha}\hat P_{\beta}|C^{k}\rangle, 
	k=1,\ldots,2^q\}\;\; .
	\label{8}
\end{equation}
The initial space after the error is given by the direct sum of all the above
subspaces, ${\cal H} = \sum_{\alpha\beta}\oplus {\cal H}_{\alpha\beta}$.  Each
time we perform an overlap and obtain a zero result, the state space $\cal H$ 
reduces in dimension, eliminating that subspace as containing the state after
the error. Eventually, one of these overlap measurements will give a positive 
result which is mathematically equivalent to projecting onto the corresponding 
subspace. The state after this projection is then given by the mixture 
$\hat\eta_f = \sum_n |\psi_{nProj_{\alpha \beta}}\rangle \bra{\psi_{nProj_{\alpha \beta}}}$, where
\begin{equation}
	|\psi_{nProj_{\alpha \beta}}\rangle = \sum_{kl}\sum_{\gamma\delta} x_n^{\gamma\delta} \hat A_{\alpha}\hat P_{\beta}
	|C^{k}\rangle\langle C^{k}| \hat P_{\beta}\hat A_{\alpha}\hat A_{\gamma}\hat P_{\delta}
	|C^l\rangle c_l\;\; .
	\label{9}
\end{equation}
The successful projection will effectively take us to the state generated by 
a superposition of certain types of error. One might expect that to 
distinguish between various errors the different subspaces ${\cal H}_{\alpha\beta}$ would have to be orthogonal. However, we will show that this is not, in fact, necessary.
  
After having projected onto the subspace ${\cal H}_{\alpha\beta}$ we now 
have to correct the corresponding error by applying the operator
$\hat P_{\beta}\hat A_{\alpha}$ onto $|\psi_{Proj_{\alpha \beta}}\rangle$,
since $\hat P_{\beta}\hat A_{\alpha}\hat A_{\alpha}\hat P_{\beta}=\hat I$. 
In order to correct for the error, the resulting state has to be proportional 
to the initial state of code--words in $|\psi_i\rangle$. 
This leads to the condition
\begin{equation}
	\sum_{kl} \sum_{\gamma\delta} x_n^{\gamma\delta} |C^k\rangle\langle C^k| \hat P_{\beta}
	\hat A_{\alpha}\hat A_{\gamma}\hat P_{\delta}|C^l\rangle c_l 
	=
	z^{\alpha\beta n} \sum_m c_m|C^m\rangle\;\; .
	\label{10}	
\end{equation}
where $z^{\alpha\beta n}$ is an arbitrary complex number. 
Now we use the fact that all code words are mutually orthogonal, {\it i.e.}
$\langle C^k|C^l\rangle = \delta_{kl}$, to obtain that  
\begin{equation}
	\sum_{l} \sum_{\gamma\delta} c_l x_n^{\gamma\delta} \langle C^{k}| \hat P_{\beta}\hat A_{\alpha}
	\hat A_{\gamma}\hat P_{\delta}|C^l\rangle
	 = z^{\alpha\beta n} c_k
	\label{11}
\end{equation}
for all $k$ and arbitrary $c_k$. This can be written in matrix form
as
\begin{equation}
	{\bf F}^{\alpha\beta n} {\bf c} = z^{\alpha\beta n} 
        {\bf c}\;\; ,
	\label{12}
\end{equation}
where the elements of the matrix ${\bf F}$ are given by
\begin{equation}
	F_{kl}^{\alpha\beta n}  := \sum_{\gamma\delta} x_n^{\gamma\delta} \langle C^{k} | \hat P_{\beta}\hat A_{\alpha}
            \hat A_{\gamma}\hat P_{\delta}|C^l\rangle\;\; .
	\label{13}
\end{equation}
As eq. (\ref{12}) is valid for all ${\bf c}$ it follows that
\begin{equation}
	\forall \;\; k, l, \;\;\;\;\;\; F_{kl}^{\alpha\beta n} = 
        z^{\alpha\beta n} \delta_{kl}\;\; .
	\label{14}
\end{equation}
However, we do not know the form of $x_n^{\gamma\delta}$'s as we 
have no information about the state of the environment. Therefore, for the 
above to be satisfied for {\it any} form of $x$'s we need each individual term in eq. (\ref{13}) to satisfy

\begin{equation}
\langle C^{k}| \hat P_{\beta}\hat A_{\alpha}
	\hat A_{\gamma}\hat P_{\delta}|C^l\rangle
	 = y^{\alpha\beta\gamma\delta} \delta_{kl}
	\label{14a}
\end{equation}
where $y^{\alpha\beta\gamma\delta}$ is {\em any} complex number. From 
eqs. (\ref{13},\ref{14},\ref{14a}) we see that the numbers $x$, $y$, and $z$ are 
related through
\begin{eqnarray}
\sum_{\gamma\delta} x_n^{\gamma\delta} y^{\alpha\beta\gamma\delta} = z^{\alpha\beta n}   \,\, .  
\end{eqnarray}

Eq. (\ref{14a}) is the main result in this section, and gives a general, and in 
fact the {\it only}, constraint on the construction of code--words, 
which may then be used for encoding purposes. If we wish to correct for up to $d$ errors, we have to impose a further constraint on the subscripts $\alpha, \beta, \gamma, \mbox{and}\; \delta$; namely, $\mbox{wt}(\mbox{supp}(\alpha) \cup \mbox{supp}(\beta)), \; \mbox{wt}(\mbox{supp}(\gamma) \cup \mbox{supp}(\delta)) \le d$,where supp($x$) denotes the set of locations where the n--tuple $x$ is different from zero and wt($x$) is the Hamming weight \cite{Pless1}, {\it i.e.} the number of digits in $x$ different from zero. This constraint on the indices of errors simply ensures that they do not contain more than $d$ logical `$1$'s 
altogether, which is, in fact, equivalent to no more than $d$ errors occurring 
during the process. 

We emphasise that these conditions are the most general possible,
and they in particular generalise the conditions in \cite{Ekert1}.  
By substituting $z^{\alpha \beta \gamma \delta}={\delta}_{\alpha 
\beta}{\delta}_{\gamma \delta}$ in eq.(\ref{14a}), we obtain the conditions
\begin{equation}
	 \langle C^{k}| \hat P_{\beta}\hat A_{\alpha}\hat A_{\gamma}\hat 
          P_{\delta}|C^l\rangle
	= \delta_{\beta\delta}\delta_{\alpha\gamma} \delta_{kl}	
	\label{15}
\end{equation}
given in \cite{Ekert1}.  These are therefore seen only as a special
case of the general result in eq. (\ref{14}).

Knill and Laflamme, who arrive at the same conditions as in eq. (\ref{14a}) 
\cite{Knill1}, give no example of a code that violates the conditions in eq. (\ref{10}) 
but satisfies those of eq. (\ref{14a}). Such a code is given by Plenio 
et al \cite{Plenio1}, which by violating the conditions  given in eq. (\ref{15}), 
explicitly shows that they are {\it not} necessary, but merely sufficient.

\section{Local Error Correction Preserves Correlations}

\subsection{Theoretical Considerations}

Imagine two initially entangled quantum systems $A$ and $B$ distributed
between two spatially separated parties. Let,
for the sake of simplicity, both $A$ and $B$ be two spin--$\frac{1}{2}$ 
particles in the initial EPR--like state 
\begin{equation}
\ket{\psi_{A+B}} = \alpha \ket{0}_A\ket{1}_B+\beta \ket{1}_A\ket{0}_B
\label{16}
\end{equation}
where the first ket describes the system $A$ and the second the system $B$.
Let both Alice particles be encoded locally (i.e. adding locally a 
certain number of auxiliary qubits and performing local unitary transformations 
to  encode) in order to protect their own qubit against a desired number of 
errors. We suppose that they both use the same coding, with the code--words
denoted by $\ket{C^0}$ and $\ket{C^1}$. After the encoding, the state is
therefore
\begin{equation}
\ket{\Psi_{A+B}} = \alpha \ket{C^0}_A\ket{C^1}_B + \beta\ket{C^1}_A\ket{C^0}_B \;\; .
\label{17}
\end{equation}
Notice that the entanglement between the systems $A$ and $B$ is not changed
by the encoding procedure, since local unitary operations do not change the 
spectrum of the reduced density matrices.

Let this state now be corrupted by errors, $\hat E$, which are local in nature, 
after which we perform the above described projections in eq. (\ref{8}) 
to obtain

\begin{equation}
\ket{\Psi_{A+B}}' = \alpha\, \hat E_i\,\ket{C^0}_A\, \hat E_j\,\ket{C^1}_B + \beta\, 
\hat E_i \ket{C^1}_A\, \hat E_j \,\ket{C^0}_B \;\; .
\label{18}
\end{equation}
We wish to show that the error does not change the value of the entanglement.
For this we compute $A$'s reduced density matrix:
\begin{eqnarray}
\hat \rho_{A} & = & \mbox{tr}_B (\ket{\Psi_{A+B}}' \bra{\Psi_{A+B}}')\nonumber \\ 
               & = & \bra{C^0}_B(\ket{\Psi_{A+B}}' \bra{\Psi_{A+B}}')\ket{C^0}_B+
\bra{C^1}_B(\ket{\Psi_{A+B}}' \bra{\Psi_{A+B}}')\ket{C^1}_B\nonumber \\
               & = & \bra{C^0}_B\, \hat E_j\ket{C^0}_B\,\{\,|\alpha|^2 \hat E_i 
\,\ket{C^0}_A\bra{C^0}_A\, \hat E_i + |\beta|^2 \, \hat E_i \,\ket{C^1}_A\bra{C^1}_A\, 
\hat E_i\,\} \; .
\label{19}
\end{eqnarray} 
which obviously has the same entropy as the original state in eqs. (\ref{15},\ref{16}) 
and eq. (\ref{17}). In the above derivation we used the 
relations in eq. (\ref{14a}) such that
\begin{eqnarray}
\bra{C^0}_B \hat E_i \, \hat E_j \ket{C^0}_B & = & \bra{C^1}_B \hat E_i \, \hat E_j 
\ket{C^1}_B\\
\bra{C^0}_B \hat E_i \, \hat E_j \ket{C^1}_B & = & 0  \; . 
\end{eqnarray}
Thus the entropy of the reduced density matrices of the initial pair of 
encoded systems, and of the systems after undergoing errors are both the same, 
indicating that the correlations and thus the entanglement do not change 
during the above described process.  By a process of introducing more local
degrees of freedom into the problem, we are able to maintain non-local 
quantum correlations. So, in fact, this process does also involve discarding
information, but is different to the post selection previously described. 
This is so because all the error correcting particles are introduced 
locally, and do not form a part of the original ensemble.

\subsection{Example With Cavities}

We now present a simple example of how to locally preserve 
entanglement between two cavities in the state
\begin{equation}
\alpha \ket{0}_A\ket{1}_B + \beta \ket{1}_A\ket{0}_B
\label{ec1}
\end{equation}
against a single amplitude error (action of $\hat\sigma_x$ Pauli operator)
on either cavity. For this purpose we {\it locally} introduce a pair of atoms 
to each cavity, all of which are in the ground state. These atoms interact 
identically with their respective cavities. We also allow errors 
to happen to the atoms, as long as there is no more than one error on 
either side, $A$ or $B$. We would like to implement the following interaction 
in order to encode the state against an amplitude error \cite{Steane2} (four 
additional atoms for each cavity are needed to correct against a general type 
of single error \cite{Laflamme1})
\begin{eqnarray}
\ket{0}\ket{g}_1\ket{g}_2 & \longrightarrow & \ket{0}\ket{g}_1\ket{g}_2 \\
\ket{1}\ket{g}_1\ket{g}_2 & \longrightarrow & \ket{1}\ket{e}_1\ket{e}_2
\label{ec2}
\end{eqnarray}
This is, in fact, an action of two `Control--Nots' \cite{Adriano}, with the
control bit being the state of the cavity and the target
bits being the atoms $1$ and $2$. We therefore perform identical
interactions on both cavities and  their atoms.  This is shown
schematically in Fig.2.
The state of the whole system (`$2$ cavities + $4$ atoms') 
will be after the encoding procedure,
\begin{equation}  
\alpha \ket{0}_A\ket{g,g}_A \, \ket{1}_B\ket{e,e}_B +
\beta \ket{1}_A\ket{e,e}_A \, \ket{0}_B\ket{g,g}_B
\label{ec3}
\end{equation}
So all we need to know is how to implement a Control--Not 
operation between the cavity and one atom. This is done in
the following way \cite{Raimond1}. Let the atom be sent
through the cavity, which in our case contains either
one or no photons, interacting resonantly with the field.
Let us in addition have a `classical' light source (a laser)
resonant with the dressed atom-field transition
$\ket{1}\ket{g} \longrightarrow \ket{1}\ket{e}$. Due to 
the vacuum Rabi splitting this will not be resonant with
$\ket{0}\ket{g} \longrightarrow \ket{0}\ket{e}$ which is
precisely what we need. In this way the initial `cavity+
atom' state undergoes evolution of the form
\begin{equation}
(\alpha \ket{0} + \beta \ket{1})\ket{g} \longrightarrow  
\alpha \ket{0}\ket{g} + \beta \ket{1}\ket{e}
\label{ec4}
\end{equation}
which is a Control--Not gate. By repeated action of
this gate we can create the state in eq. (\ref{ec3}).
Then if a single amplitude error occurs on either side
(e.g. a spontaneous decay of the field)
we can correct it by applying a unitary operation to the
cavities to restore the original state, depending on the 
state of the four atoms \cite{Steane2}.

Let us give a simple example of how this would work. Suppose 
that only the cavity $A$, after encoding, undergos an amplitude error
resulting in, after a small rearrangement, the joint `cavities + atoms+environment' state of the 
form (eq. (\ref{7a}))

\begin{eqnarray}  
&   & (\alpha \ket{0}_A\ket{1}_B \ket{g,g}_A \, \ket{e,e}_B +
\beta \ket{1}_A\ket{0}_B \ket{e,e}_A \, \ket{g,g}_B) \ket{R_0} \nonumber \\
& + &  (\alpha \ket{1}_A\ket{1}_B \ket{g,g}_A \, \ket{e,e}_B +
\beta \ket{0}_A\ket{0}_B \ket{e,e}_A \, \ket{g,g}_B) \ket{R_1}\;\; .
\label{ec5}
\end{eqnarray}
To recover the original state we first have to decode the above state. 
This is just the inverse of encoding, i.e. we apply two Control-Nots 
as described above, resulting in the state

\begin{eqnarray}  
&   & (\alpha \ket{0}_A\ket{1}_B  +
\beta \ket{1}_A\ket{0}_B ) \ket{g,g}_A \, \ket{g,g}_B) \ket{R_0} \nonumber \\
& + &  (\alpha \ket{1}_A\ket{1}_B +
\beta \ket{0}_A\ket{0}_B) \ket{e,e}_A \, \ket{g,g}_B) \ket{R_1}\;\; .
\label{ec6}
\end{eqnarray}

In the second step we can make a measurement on the atoms and depending on
the outcome apply an appropriate unitary transformation to the cavities. In 
this case we only have to consider cavity $A$:
if both of the atoms are in the ground state then we do nothing because the 
joint-cavity state remains unchanged, whereas if
both of the atoms are excited we apply a NOT operation to cavity $A$. 
This we do in a fashion similar to performing Control--Not. We could, for example, send an excited atom throught the cavity and tune the external 
laser to the dressed transition 
$\ket{0}\ket{e}\longleftrightarrow \ket{1}\ket{e}$.  In this way 
we recover the state in eq. (\ref{ec1}). We emphasise that 
the form in eq. (\ref{ec6}) is incomplete since the terms arising from all 
the other amplitude errors are missing (corresponding to the cavity $B$ and the atoms); however, it can easily be checked that the above scheme would also accomodate for this.

\section{Conclusion}

In this paper we presented simple models to demonstrate that correlations 
cannot be increased by any form of local complete measurement.
The consequence of this is that any purification procedure has 
to represent a post-selection of the original ensemble to be purified. Classical
communication is an essential precursor to the post-selection procedure --- we
cannot post-select without classical communication, but the post-selection
procedure is necessary to prepare the maximally entangled subset. We
then presented two general proofs of this fact. Additionally, we derived general
conditions which have to be satisfied by quantum error correction codes,
which can be used to protect a state against an arbitrary number of errors. 
We then showed that we can locally `protect' 
the entanglement by standard quantum error correction schemes, such that the 
correlations (and therefore the entanglement) are preserved under any type of 
complete measurement, which can be viewed as an error in this context. We gave
a simple example of how to encode two cavities against a single amplitude 
error. Thus, local error correction can protect nonlocal features of entangled 
quantum systems, which otherwise cannot be increased by any type of local 
actions which exclude classical communication and post-selection.

\newpage
{\bf Acknowledgements}\\[.25cm]
The authors thank P.L.Knight, and V.V. and M.B.P thank A. Barenco, A. Ekert and 
C. Macchiavello, for useful discussions on the subject of this paper.
This work was supported by the European Community,
the UK Engineering and Physical Sciences Research 
Council and by a Feodor-Lynen grant of the Alexander von Humboldt 
foundation.

\vskip 1cm
\begin{center}
{\bf FIGURE CAPTIONS}
\end{center}

Figure 1: 
The experimental setup for local interactions: two cavities are
initially entangled in a state of the form $\ket{\Psi}_{AB} = \alpha\ket{n}_A\ket{m}_B + \beta\ket{n^{\prime}}_A\ket{m^{\prime}}_B$, 
and atoms are sent through cavity $A$ only.\\[1.5cm]

Figure 2: 
The encoding network for protecting against amplitude errors is shown
in the upper diagram: the encircled cross denotes a NOT operation while 
a dot denotes a control bit, together making a Control--Not operation. 
The atoms are initially in their ground states, and the order in which 
the gates are executed is irrelevant.  The lower diagram gives a truth
table for the Control--Not operation; here, `C' and `T' represent 
{\it control} and {\it target} bits respectively.

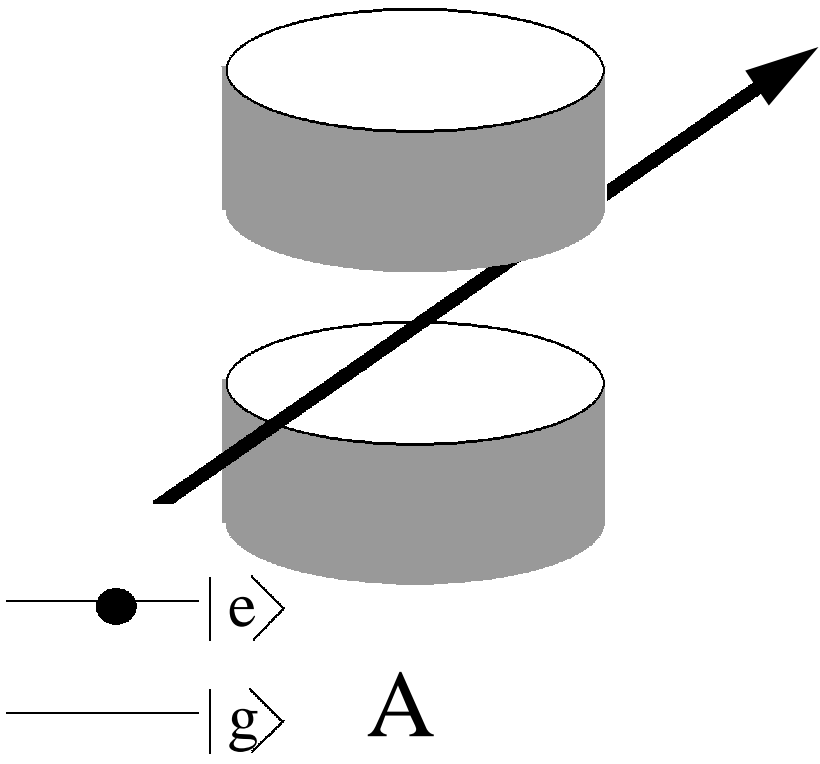
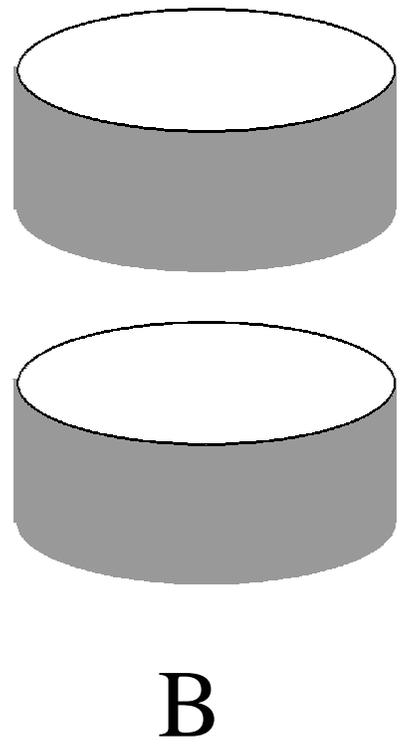
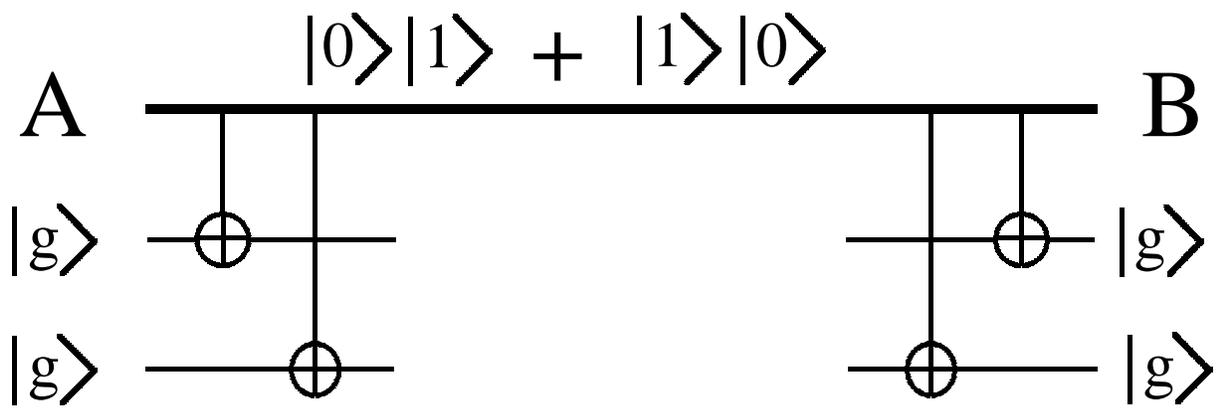
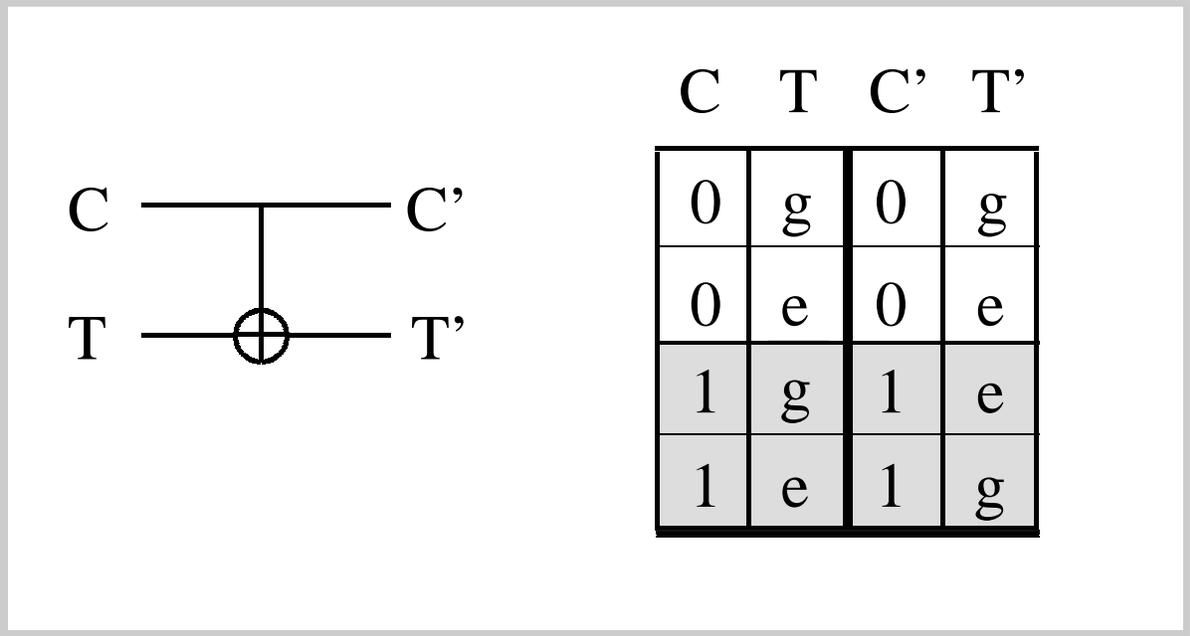
\end{document}